\documentclass[12pt]{iopart}

\usepackage{color} 

\usepackage{graphicx}
\usepackage{epstopdf}

\begin{document}


\title[Chronic neural probe for recording of SUA, MUA, and LFP activity]{Chronic neural probe for simultaneous recording of single-unit, multi-unit, and local field potential activity from multiple brain sites}

\author{F~Pothof$^1$, L~Bonini$^2$, M~Lanzilotto$^3$, A~Livi$^3$, L~Fogassi$^3$, G~A~Orban$^3$, O~Paul$^{1,4}$ and P~Ruther$^{1,4}$}
\address{$^1$ Department of Microsystems Engineering (IMTEK), University of Freiburg, Freiburg, DE}
\address{$^2$ Istituto Italiano di Tecnologia (IIT), Brain Center for Social and Motor Cognition, Parma, IT}
\address{$^3$ Dipartimento di Neuroscienze, Universit\`{a} degli studi di Parma, Parma, IT}
\address{$^4$ BrainLinks-BrainTools, University of Freiburg, Freiburg, DE}
\ead{frederick.pothof@imtek.uni-freiburg.de}

\begin{abstract}
\textit{Objective.} Drug resistant focal epilepsy can be treated by resecting the epileptic focus requiring a precise focus localization using stereo\-electro\-encephalo\-graphy (SEEG) probes. As commercial SEEG probes offer only a limited spatial resolution, probes of higher channel count and design freedom enabling the incorporation of macro and microelectrodes would help increasing spatial resolution and thus open new perspectives for investigating mechanisms underlying focal epilepsy and its treatment. This work describes a new fabrication process for SEEG probes with materials and dimensions similar to clinical probes enabling recording single neuron activity at high spatial resolution.
\textit{Approach.} Polyimide is used as a biocompatible flexible substrate into which platinum electrodes and leads are integrated with a minimal feature size of 5\,$\mu$m. The polyimide foils are rolled into the cylindrical probe shape at a diameter of 0.8\,mm. The resulting probe features match those of clinically approved devices. Tests in saline solution confirmed the probe stability and functionality. Probes were implanted into the brain of one monkey (Macaca mulatta), trained to perform different motor tasks. Suitable configurations including up to 128 electrode sites allow the recording of task-related neuronal signals.
\textit{Main results.} Probes with 32 and 64 electrode sites were implanted in the posterior parietal cortex. Local field potentials and multi-unit activity were recorded as early as one hour after implantation. Stable single-unit activity was achieved for up to 26 days after implantation of a 64-channel probe. All recorded signals showed modulation during task execution.
\textit{Significance.} With the novel probes it is possible to record stable biologically relevant data over a time span exceeding the usual time needed for epileptic focus localization in human patients. This is the first time that single units are recorded along cylindrical polyimide probes chronically implanted 22\,mm deep into the brain of a monkey, which suggests the potential usefulness of this probe for human applications.
\end{abstract}


\submitto{\JNE}

\maketitle

\section{Introduction}

Neurological disorders, such as epilepsy, Parkinson's disease, and multiple sclerosis, constitute the cause of approximately 12\% of all deaths worldwide. Furthermore, it has been estimated that epilepsy alone accounts for about 7.9\% of the total disability-adjusted life years (DALY) caused by all neurological disorders \cite{WHO2006}. Already today, a large number of patients suffering from these diseases greatly benefit from advances in neurophysiological techniques \cite{Scarabin2012, Cossu2005, Rodriguez2005, Wishart2003}. This includes among others the monitoring of local field potentials (LFP) prior to the surgical treatment of focal epilepsy \cite{Scarabin2012, Cossu2005}, or  deep brain stimulation in the case of Parkinson's disease \cite{Rodriguez2005}, multiple sclerosis related tremor \cite{Wishart2003}, treatment of drug resistant depression \cite{Mayberg2005}, and the Tourette's syndrome \cite{Houeto2005}. In the case of drug resistant epilepsy, i.e. about 22.5\% of all epilepsy patients \cite{Picot2008}, originating from a single area in the brain (focal epilepsy) a more invasive method of treatment can be chosen, namely, the resection of the specific brain region. For this purpose, surgeons have to precisely localize the focus. In case the epileptic focus is relatively superficial, electrocorticography (ECoG) is used for localization \cite{Truccolo2011}. In contrast, when the  epileptic focus is most likely positioned in deeper cortical regions, penetrating depth probes are the tool of choice to guarantee a better spatial resolution \cite{Schulze2014} than using surface probes. As a large part of the brain has to be electrophysiologically explored in the clinical diagnosis of focal epilepsy, several so called stereoelectroencephalography (SEEG) probes hosting multiple recording sites along the same shank are implanted into the cortex in order to precisely localize the position of the epileptic focus.

\begin{figure}[tbh]
	\centering
	\includegraphics[width=8cm]{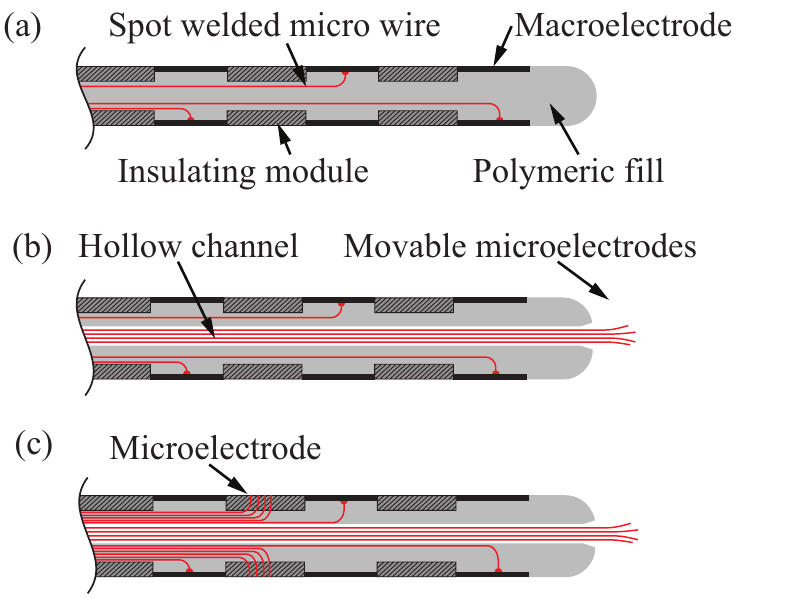}
	\caption{Schematics of approved depth probe approaches. Cross-sections of (a) a conventional SEEG probe comprising hollow metal cylinders spot-welded to micro wires and insulating modules, (b) an advanced SEEG probe with larger diameter containing additional movable Behnke-Fried microelectrodes integrated in the hollow probe body (on the basis of \cite{Misra2014}), (c) and micro contacts are positioned between the macro contacts (on the basis of \cite{Worrell2008}).}
	\label{fig:overview}
\end{figure}

Clinically approved SEEG probes usually comprise electrodes made from hollow platinum-iridium (PtIr) modules and insulator modules alternating in cylindrical assemblies, as schematically shown in Figure\,\ref{fig:overview}(a). Insulated micro wires are attached on the inside of the electrodes by spot-welding followed by the manual assembly of electrodes and insulating modules, and mechanically stabilized by a subsequent polymeric filling of the hollow probe body. Although this approach has been highly optimized over the past decades, the number of electrodes is inherently limited by the small space available in the centre of the probe for gathering and guiding the interconnecting wires of the individual electrode cylinders. Probe diameters between 0.8\,mm and 1.27\,mm (with some variation among the manufacturers and types of probes) carrying up to 18 electrodes have been clinically approved and are commercially available \cite{Avanzini2016}. The lengths of the probes vary among pre-surgically defined trajectories and patients, but usually do not exceed 8\,cm for human applications. The duration of such implantations is typically about 10 days.

A modification of the clinically applied SEEG probes with a slightly increased diameter of 1.27\,mm  comprises a hollow channel along the probe centre \cite{Fried1997}. As illustrated in Figure\,\ref{fig:overview}(b), a bundle of eight insulated micro wires with a diameter of 40\,$\mu$m each is guided through this channel. This bundle is movable, enabling a subsequent deeper penetration into the surrounding brain tissue once the SEEG probe has been put in place. While the cylindrical macro contacts are again used to record LFPs along the probe shank, these so-called Behnke-Fried microelectrodes, can be further advanced into the neural tissue around the probe tip, enabling the measurement of single-unit activity (SUA), multi-unit activity (MUA), and LFPs in the medial surface of the hemispheres \cite{Fried1997}. These SEEG probes with macroelectrodes and movable microelectrodes are clinically approved and have been used in various studies on brain physiology \cite{Misra2014, Fried1997, Kawasaki2001, Mukamel2010, Hefft2013, Alvarado2013}. They have also been successfully applied in recording SUA deep within the cortex during an epileptic seizure \cite{Alvarado2013} greatly advancing the knowledge about the mechanisms underlying focal epilepsy. However, the fact that microelectrodes protruding from the probe tip are applied in these modified SEEG probes clearly restricts their capability in recording SUA to the medial side of the hemispheres. A variation of this approach, first proposed by Worrell et al. \cite{Worrell2008,Gompel2008},  includes 18 additional microelectrodes along the shank, and constitutes a first step towards neural recording with high-density microelectrodes integrated along the length of cylindrical neural probes \{Figure\,\ref{fig:overview}(c)\}. Furthermore, Ulbert et al. have shown that SUA can be successfully recorded acutely in humans from up to 24 microelectrodes positioned along the shank of semi-cylindrical neural depth probes \cite{Ulbert2004,Csercsa2010}, motivating to extend these recordings into the domain of chronic recordings with an increased channel count.

\begin{figure}[tbh]
	\centering
	\includegraphics[width=8cm]{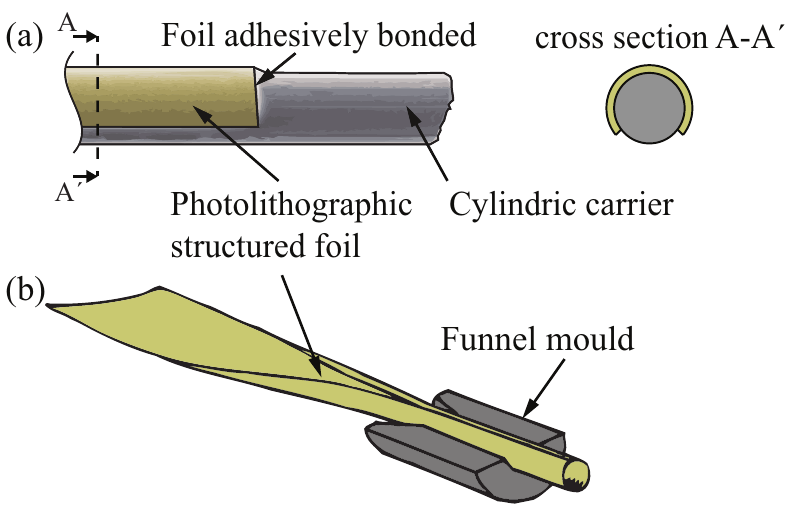}
	\caption{Schematics of alternative neural probe approaches based on polymer foils with integrated metallization realized using microfabrication technologies (a) attached to a carrier rod, with cross-section along the dashed line (adapted from \cite{Fomani2011a}) and (b) transferred into the cylindrical probe shape using a dedicated mould (adapted from \cite{Puije1989}, for illustration purposes only the lower half of the mould is shown).}
	\label{fig:overviewb}
\end{figure}

Alternative approaches for fabricating depth probes that circumvent the labour intensive assembly of individual metal cylinders connected to micro wires are based on polymer foils with integrated thin-film metal structures realized using microfabrication technologies \cite{Fomani2011a, Mercanzini2010, Martens2008}. As demonstrated by Fomani et al. \cite{Fomani2011a} and similarly proposed in a patent application by A. Mercanzini and P. Renaud \cite{Mercanzini2010}, these foils can be wrapped around a cylindrical carrier to which  they are adhesively bonded, as schematically shown in Figure\,\ref{fig:overviewb}(a). The assembly process of the planar electrode array onto its carrier applies a thermoplastic film patterned by laser micromachining and a two-piece mould in which both components are manually aligned. As indicated by Figure\,\ref{fig:overviewb}(a), the assembly process does not allow the foil to cover the entire carrier circumference and thus limits the number of interconnecting leads that can be integrated. Nonetheless, probes with up to 32 channels and an outer diameter as small as 0.75\,mm have been realized using this approach \cite{Fomani2011a}. To our best knowledge, only in vitro LFP recordings and electrical stimulation experiments using explanted mice brains have been performed using these probes \cite{Fomani2011a}. Already in 1989, van der Puije et al. proposed a different probe shaping process for cochlear implants that applies a mould with a lead-in funnel \cite{Puije1989}. By introducing the planar probe foil into the mould, the foil is rolled into the cylindrical probe shape, as illustrated in Figure\,\ref{fig:overviewb}(b). This shape is preserved by filling the inner probe volume with silicone, while the rolled foil is still inside the mould. Based on the images in Ref.\,\cite{Puije1989} it is however unclear whether the rolled ends of the foil overlap or leave a gap, presumably resulting in a non-cylindrical probe geometry. 

This paper presents a chronic depth probe combining macroelectrodes similar to those of conventional SEEG probes used for LFP recordings in epilepsy diagnostics and microelectrodes with the capability of SUA recording along the entire cylindrical probe shank. In contrast to the Behnke-Fried approach with protruding micro wire electrodes \cite{Fried1997}, the access to SUA is no longer limited to the probe tip.  Similarly to previous depth probes \cite{Fomani2011a}, cochlear implants \cite{Puije1989}, probe cables \cite{Hetke1990}, and ECoG microelectrode arrays \cite{Stieglitz2000} designed for non-clinical research, the SEEG probe developed in this study is based on a polymer foil with embedded high-resolution electrical leads. In contrast to earlier cylindrical neural probes \cite{Fomani2011a} and comparable to cuff electrodes \cite{Rodriguez2000, Plachta2014}, the lateral edges of the foil overlap by at least one revolution enabling a pronounced increase in the channel count. This would allow to generate highly resolved four-dimensional maps of human cortical processing with an even smaller number of patients \cite{Avanzini2016}. As a consequence, a 128-channel prototype is successfully demonstrated in this study. Similar to the work of van der Puije et al. \cite{Puije1989}, a dedicated mould serves the definition of the probe geometry during the probe shaping process. This process relies on an annealing step at an increased temperature and a photo-lithographically patterned dry adhesive \cite{Pothof2014}. In contrast to Fomani et al. \cite{Fomani2011a}, a probe carrier is avoided as the assembly process provides a self-supporting probe body stabilized by the dry adhesive. Aside from the probe fabrication and bench testing, we present data gathered during the in vivo validation of SEEG probes with 32 and 64 channels and an outer diameter of 0.8\,mm by recording of LFP, MUA, and SUA throughout the hemisphere of a non-human primate.

\section{Materials and methods}

\subsection{Depth probe design}

\begin{figure}[tbh]
	\centering
	\includegraphics[width=15.7cm]{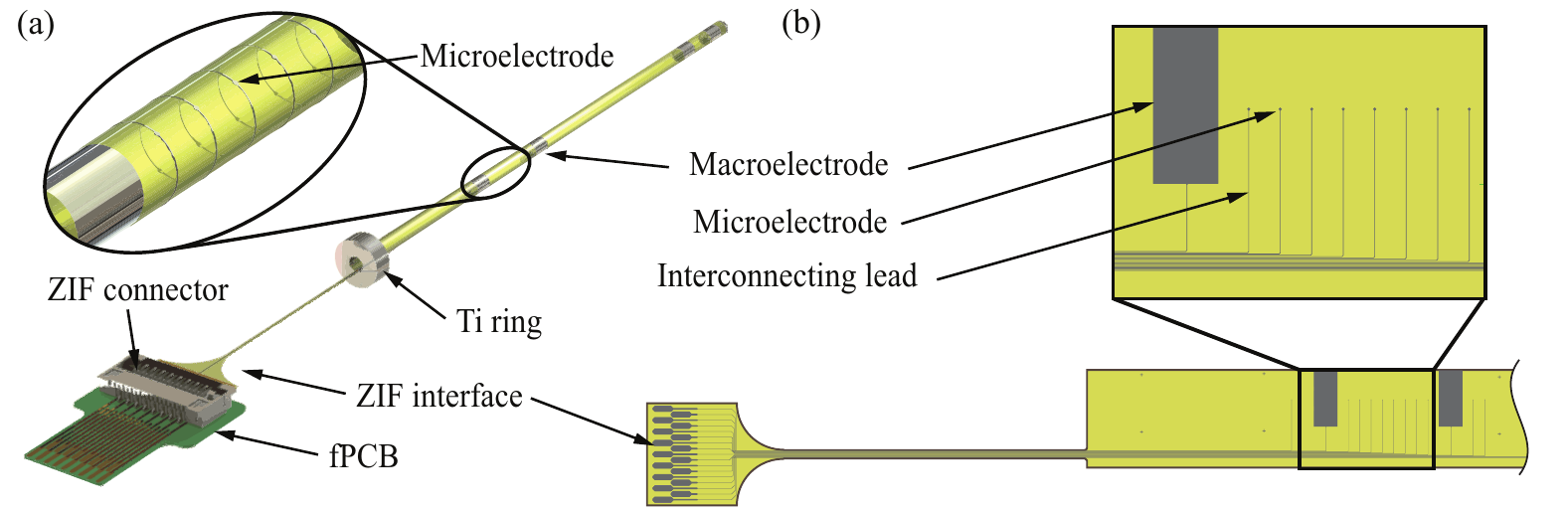}
	\caption{(a) Computer model of the polyimide cylinder filled with epoxy exhibiting four macroelectrodes with microelectrodes positioned in between (detail). A titanium ring is glued to the proximal end to control the implantation depth. The planar end of the polyimide foil is connected to a commercial zero-insertion-force (ZIF) connector allowing probe interfacing via regular flexible printed circuit board (fPCB) technology. (b) Planar foil layout.}
	\label{fig:Probe}
\end{figure}

The cylindrical depth probe developed in this study is schematically shown in Figure\,\ref{fig:Probe}(a) as a simplified computer model. It is made from a stack of two polyimide (PI) films (each 5\,$\mu$m thick) with electrodes and interconnecting lines sandwiched in-between. The probe metallization is realized in platinum (Pt) to define macroelectrodes and microelectrodes located at pre-operatively defined positions aiming for the recording of LFP and SUA in specifically targeted brain regions. The cylindrical probe is electrically interconnected to the external instrumentation via a planar section of the PI structure that fits into a high-density zero-insertion-force (ZIF) connector. For probe implantation, a hollow screw with an inner diameter slightly larger than the probe diameter is used,  similar to the clinical application of conventional SEEG probes in epilepsy diagnostics. It defines the probe orientation during implantation and serves as a mechanical fixture. In order to precisely define the implantation depth, a titanium (Ti) ring, as shown in Figure\,\ref{fig:Probe}(a), is attached to the proximal end of the depth probe. This ring is in planar contact with the screw head after probe insertion into the brain tissue. It may be fixed to the skull screw either by a drop of silicone, bone cement, or a screw nut. 

\subsection{Probe fabrication}

As described in further detail below, the cylindrical probe originates from a planar PI-based foil comprising a thin-film metallization. A simplified layout of this foil is illustrated in Figure\,\ref{fig:Probe}(b). It is produced on a wafer using microfabrication technologies defining the geometry of the different types of electrodes, their interconnecting lines to the ZIF interface, as well as the outer geometry of the foil. The PI foil is subsequently shaped into its intended cylindrical probe geometry using an approach similar to the idea shown in Figure\,\ref{fig:overviewb}(b) and the fabrication process of cuff electrodes detailed elsewhere \cite{Plachta2014}. Once the cylindrical probe shape has been obtained, the hollow probe body is mechanically stabilized by a polymer. 

\begin{figure}[tbh]
	\centering
	\includegraphics[width=8cm]{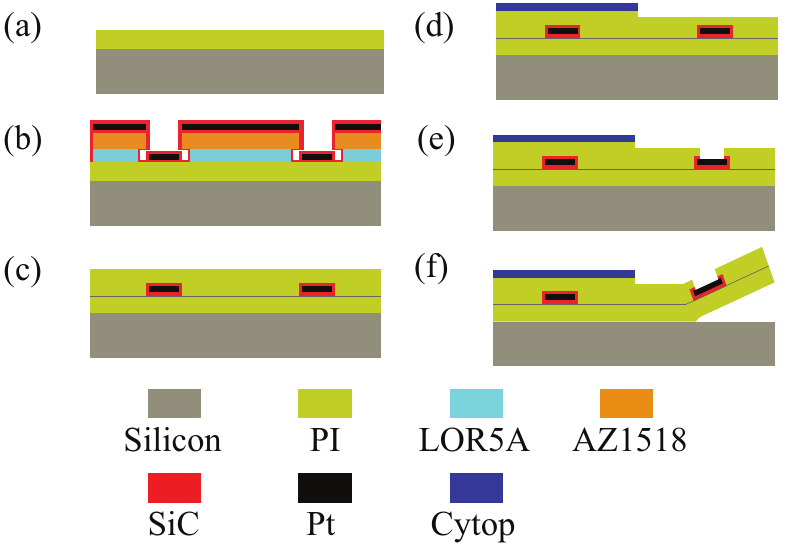}
	\caption{Fabrication of the PI foils relies on (a) spin coating of PI, (b) deposition and patterning of probe metallization, silicon carbide (SiC) and diamond like carbon (DLC) layers for improved adhesion, (c) spin coating of the second PI layer, (d) deposition and patterning of Cytop, (e) opening of electrode sites and ZIF contact pads using RIE, and (f) peeling of the foils using tweezers. Schematics are not to scale.}
	\label{fig:Process}
\end{figure}

The fabrication sequence to realize the planar PI layer stack of the cylindrical depth probe is summarized in Figure\,\ref{fig:Process}. It combines processes that have been widely used at our department to fabricate, among other structures, cuff electrodes \cite{Rodriguez2000,Plachta2014}, highly-flexible interfaces for silicon-based neural probes \cite{Kisban2007,Herwik2009,Seidl2011}, electrocorticography (ECoG) electrode arrays \cite{Rubehn2009}, and optical probes tailored for optogenetic research \cite{Schwaerzle2016}. These process steps are extended in the case of the cylindrical probes by an additional process step that defines a patterned dry adhesive layer needed for the subsequent probe shaping.

Standard silicon wafers (100\,mm diameter, thickness 525\,$\mu$m) are cleaned with acetone, isopropanol, Piranha solution and deionised water. This is followed by the removal of the native silicon oxide layer in 1\% hydrofluoric acid. As a result the adhesion of the subsequently spin-coated 5-$\mu$m-thin polyimide film (U-Varnish-S, UBE Industries Ltd., Tokyo, Japan) \{Figure\,\ref{fig:Process}(a)\} to silicon is sufficiently strong for all following fabrication steps and at the same time weak enough to enable the PI foil to be peeled off the silicon wafer at the end of the fabrication process \{Figure\,\ref{fig:Process}(f)\}. The photoresists LOR5A (MicroChem Corp., Westborough, MA, USA) and AZ1518 (Microchemicals GmbH, Ulm, Germany) with thicknesses of 700\,nm and 1.8\,$\mu$m, respectively, are spin coated onto the PI to generate a bilayer lift-off resist. It is patterned using UV-lithography. Developing the resist stack in tetramethylammonium hydroxide (TMAH) generates an undercut in the LOR5A of around 1\,$\mu$m underneath the AZ1518. Potential resist residues are removed in an oxygen plasma which activates the PI surface and thus improves the adhesion of the subsequently deposited layer stack of silicon carbide (SiC), platinum (Pt), SiC, and diamond like carbon (DLC) \{Figure\,\ref{fig:Process}(b)\}. As demonstrated by Ordonez et al. \cite{Ordonez2012}, SiC as well as DLC promote the adhesion between the PI and the Pt metallization. While SiC and DLC layers are deposited by plasma enhanced chemical vapour deposition (PECVD) at 100$^\circ$C, the 300\,nm-thin Pt layer is sputter deposited \{Figure\,\ref{fig:Process}(b)\}. Next, the resist stack is removed in acetone, followed by the activation of the PI and DLC surfaces using an oxygen plasma, and the deposition of a second 5-$\mu$m-thick PI layer \{Figure\,\ref{fig:Process}(c)\}.

A 700-nm-thin layer of the fluoropolymer Cytop (Asahi Glass Co., Tokyo, Japan) is then spun on and patterned by reactive ion etching (RIE) using 10\,$\mu$m AZ9260 photoresist (Microchemicals GmbH) as the masking layer \{Figure\,\ref{fig:Process}(d)\}. The Cytop serves as a thin film dry adhesive during the 3D shaping process of the cylindrical depth probes. The etching time and thus the resulting etching depth are carefully controlled, in order not to etch too deep into the underlying PI. Finally, the PI layer stack is patterned by RIE using again AZ9260 with a layer thickness of 24\,$\mu$m to define the foil outline and open the electrodes and ZIF contact pads. A sulphur hexafluoride (SF$_{6}$) plasma with oxygen admixture ensures the removal of the SiC and DLC layers at the openings to the Pt metallization \{Figure\,\ref{fig:Process}(e)\}. Finally, the batch fabricated PI foils are ready to be peeled off the silicon wafer using tweezers \{Figure\,\ref{fig:Process}(f)\}.

\begin{figure}[tbh]
	\centering
	\includegraphics[width=8cm]{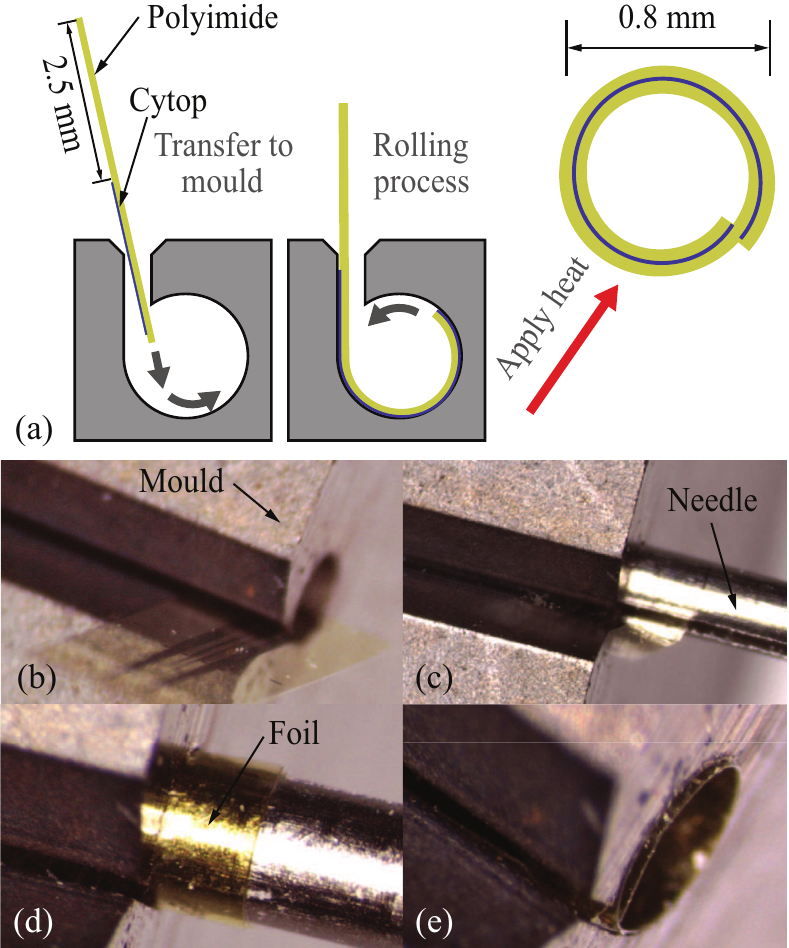}
	\caption{(a) Schematic (not to scale) and (b)-(e) micrographs taken during successive stages of the manually executed rolling process.}
	\label{fig:Rolling}
\end{figure}

The 3D probe shaping process is schematically shown in Figure\,\ref{fig:Rolling}(a). Optical micrographs taken at different stages of this shaping process are shown in Figures\,\ref{fig:Rolling}(b) to (e). First, the planar PI foil is manually inserted through the tangential entrance slit (width 80\,$\mu$m) of a cylindrical hole with a diameter corresponding to the desired probe size \{Figure\,\ref{fig:Rolling}(b)\}. The custom made mould containing the hole with entrance slit is realized using electrical discharge machining (EDM) with an accuracy and surface roughness below 1\,$\mu$m. A steel needle furnished with a 100-$\mu$m-wide slit, introduced into the needle side wall using EDM as well, is inserted axially into the mould. At this point, the foil edge has entered the needle slit, as shown in Figure\,\ref{fig:Rolling}(c). Rotating the needle inside the mould draws the PI foil along its entire length into the mould cavity. As a consequence, the PI foil is wrapped around the needle \{Figure\,\ref{fig:Rolling}(d)\}. The needle is then carefully retracted from the mould leaving the rolled foil inside \{Figure\,\ref{fig:Rolling}(e)\}. By correctly designing the foil and Cytop dimensions, it is ensured that two foil revolutions are obtained inside the mould with one full revolution of Cytop in contact with the inner surface of the polyimide foil, as illustrated in \{Figure\,\ref{fig:Rolling}(a)\}. To make sure the Cytop completely bonds to the PI at all desired areas, its glass transition temperature ($T_g \approx 110^\circ$C) is exceeded for four hours while the foil remains in the mould. To achieve at the same time a stress relaxation of the rolled PI, an anneal temperature of 290$^\circ$C is chosen. The Cytop forces the foil to retain its cylindrical shape once it is removed from the mould and prevents epoxy from leaking out laterally during the subsequent filling procedure.

\begin{figure}[tbh]
	\centering
	\includegraphics[width=8cm]{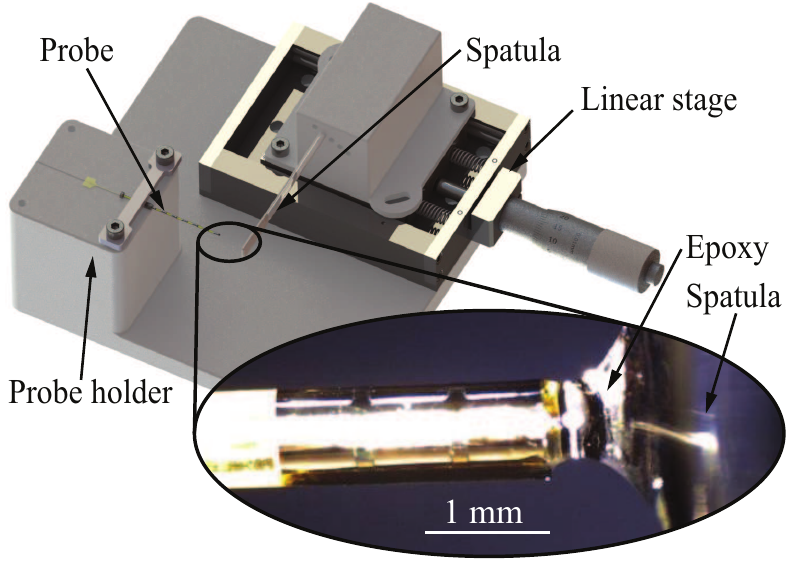}
	\caption{Computer model of a platform dedicated to filling the hollow cylindrical PI foils and (detail) micrograph of the filling process employing capillary forces.}
	\label{fig:Filling}
\end{figure}

To fill the hollow cylindrical PI foil, we rely on capillary action to draw a biocompatible epoxy (EPO-TEK\textsuperscript{\textregistered} 301-2, Epoxy Technology Inc., Billerica, MA, USA) into the cylinder \{Figure\,\ref{fig:Filling}\}. Once the epoxy reaches the opposite end of the probe cylinder, the filling process self-terminates. The epoxy is then cured at 80$^\circ$C for at least three hours to ensure complete curing. The application of a small amount of additional epoxy to the distal end of the cured probe allows a hemispherical probe tip to be realized, which facilitates probe insertion.

The electrical interfaces on the PI foil are designed to fit commercially available ZIF connectors (502078-1710 from Molex, Lisle, IL, USA). Since the metallization layer on the PI foil with its thickness of only a few hundred nanometres is prone to beeing scratched during repetitive mechanical interaction, we chose to connect the foil to a small intermediate flexible printed circuit board (fPBC). This intermediate fPCB forwards all channels from its ZIF connector (female) to a ZIF plug (male). This ZIF plug can be repeatedly disconnected from and reconnected to the corresponding counterpart on external instrumentation. Thereby  the more delicate contacts on the polyimide cable are left unharmed. The assembly of this intermediate fPCB with the probe is shown in Figure\,\ref{fig:Probe}(a). The metal layer of this intermediate ZIF plug is 35\,$\mu$m thick and thus resistant to failure during multiple insertion cycles which are required by the daily probe application in experiments lasting several weeks. To protect this intermediate ZIF from being opened accidentally during probe application, epoxy is used to fix the opening lever of the ZIF connector. This intermediate ZIF plug can be inserted into its counterpart on a rigid electrode interface board which finally makes the connection to the recording equipment.

\subsection{Characterization of the depth probe}

Before implantation, the probes were characterized with respect to their geometrical specifications, mechanical insertion stability, and electrical properties. 

As the probe diameter is most relevant for probe insertion through the guiding screw for defining the probe orientation and positioning, only tight tolerances are allowed. To inspect the diameter of the probes we used a calibrated optical microscope. 
Scanning electron microscopy (SEM) is further employed to inspect the openings of the polyimide on the electrode sites. As a mechanical test vehicle, a 0.6\% agar gel-based model of a macaque brain was used to confirm that no buckling nor fracture occurred during insertion. For the electrical characterization we used the commercial electrochemical impedance analyser CompactStat (Ivium Technologies, Eindhoven, The Netherlands).

In addition, impedance measurements at 1\,kHz were employed to validate that the influence of common sterilization procedures, i.e. autoclave sterilization at 134$^\circ$C and at 121$^\circ$C, and a STERRAD\textsuperscript{\textregistered} plasma process, on probe performance is weak. We employed four 64-channel probes for these tests and monitored the stability of the electrode impedances before and after sterilization. For comparison, a reference probe was transported and handled in exactly the same manner as the other probes, except that it did not undergo sterilization. Results are reported in Section \ref{sec:Character}.

\subsection{Neurophysiological methods: probe implantation, apparatus and behavioural paradigm}

Experiments have been performed on one adult female \itshape{Macaca mulatta}\normalfont{} (4\,kg).
Prior to the recordings, the monkey was trained to sit in a primate chair, to interact with the experimenters, and to grasp visually presented objects (a ring, a big cone, and a small cone) using the right hand contralateral to the left hemisphere to be recorded from as described earlier \cite{Bonini2014}, and was habituated to receive different types of passive sensory stimulation \cite{Rozzi2008,Maranesi2012}.

Once the training was completed, a head fixation system and a plastic recording chamber were implanted under general anesthesia (ketamine hydrochloride, 5\,mg/kg intramuscular (i.m.) and medetomidine hydrochloride, 0.1\,mg/kg i.m., repeatedly administered during the surgery). The surgical procedures were the same as previously described \cite{Bruni2015}. To implant a cylindrical probe, a hole was drilled into the skull in correspondence to the selected stereotaxic coordinates inside the recording chamber by means of a custom-made stereotaxic holder for the drill-bit, in order to maintain a correct angle. Then, after opening the dura, the guiding screw was implanted by means of the same stereotaxic holder. A rigid metallic cylindrically shaped stylet with a 0.8\,mm diameter was introduced through the guiding screw, to create a guiding channel into the brain for the subsequent insertion of the probe. A stopper was positioned and mechanically fixed at a distance of 30\,mm from the tip of the stylet corresponding to the distance between the probe tip and and the titanium ring. In the meantime, the probe was sterilized by submersing it into 100\% alcohol for 1 minute, rinsing it in saline solution at 100$^\circ$C, and then moving it to the surgery room within sterile saline solution at ambient temperature. Finally, the probe was inserted through the screw, advanced into the brain along the guiding channel, and fixed to the top of the guiding screw. The electrical PCB connector was mechanically fixed inside the chamber to prevent excessive forces from acting on the connector during free behaviour of the monkey. 

Dexamethasone and prophylactic broad-spectrum antibiotics were administered pre- and  postoperatively. Furthermore, analgesics were administered intra- and postoperatively. During all surgeries, hydration was maintained with continuous infusion of saline solution. A heating pad was used to maintain the body temperature constant. The heart rate, blood pressure, respiratory depth, and body temperature were continuously monitored. Upon recovery from anaesthesia, the animal was returned to its home cage and closely monitored. All experimental protocols complied with the European law on the humane care and use of laboratory animals (directives 86/609/EEC, 2003/65/CE, and 2010/63/EU), and are authorized by the Italian Ministry of Health (D.M. 294/2012-C, 11/12/2012) and approved by the Veterinarian Animal Care and Use Committee of the University of Parma (Prot. 78/12 17/07/2012).

During the recording sessions, the task phases were automatically controlled and monitored by a LabVIEW-based software, enabling to interrupt the trial in case the monkey broke visual fixation, made an incorrect movement, or did not respect the temporal constraints of the task. In all these cases, no reward was delivered. After correct accomplishment of a trial, the monkey was automatically rewarded with some drops of fruit juice. The electrical activity was recorded in 15 trials for each task.

\subsection{Recording techniques and neurophysiological data analysis}

In total we were able to test three cylindrical probes. Two of these probes, in the following termed P1 and P2, had 32 electrode channels. The electrophysiological signals were amplified and sampled at 40\,kHz with a 16-channel OmniPlex system (Plexon, Dallas, Texas, USA). An online spike sorting software (Plexon) enabled to monitor the quality and stability of the recordings during the recording sessions. The third probe, termed P3, had 64 channels and required an acquisition system that can digitize more channels simultaneously. For this purpose we used two 32-channel Intan amplifier boards (Intan Technologies, Los Angeles (CA), USA), controlled in parallel via the electrophysiology platform Open Ephys (http://www.open-ephys.org/). All formal signal analyses were performed off-line. Contact-sensitive devices (Crist Instruments, Hagerstown, MD, USA) were used to detect when the monkey touched the metal surface of the starting position or the target object with the hand. A switch located behind the object signalled the onset of the object-pulling phase. Transistor-transistor logic (TTL) signals were used by a LabVIEW-based software to monitor the monkey's performance and to control the presentation of the cues of the behavioural paradigm as well as the reward delivery (for details see \cite{Bonini2014}). All these signals were fed to the recording system to be stored in parallel with neuronal activity, in order to use them for aligning the activity, to construct response histograms, and to identify the epochs for statistical analysis of neuronal responses. Eye position was monitored with an eye-tracking system composed of a 50\,Hz infra red sensitive CCD video camera (F11CH4, Ganz/CBC Co., Ltd., Tokyo, Japan) and two spots of infra-red light.

All recorded data were processed off-line by means of a dedicated spike sorting software (Plexon, Dallas (TX), USA). Data were then analyzed with custom software written in MATLAB (Mathworks) and with CHRONUX (http://chronux.org/) toolboxes. For the reaching/grasping motor task, the trials were aligned with the visual presentation of objects and with the hand reaching or object pulling events. A period ranging from 2\,s before and 1.5\,s after each event was then selected for further analysis.
Regarding SUA and MUA, the wide-band activity was first high-pass filtered at 300\,Hz. Then, waveform detection was performed for each channel by employing a negative threshold corresponding to the three-fold standard deviation from the mean peak height of the signal. After artefact removal, all the detected waveforms were considered as task related MUA in case the spike rate showed some overall significant  modulation during at least one of the epochs of the behavioural task (repeated measures ANOVAs with p $<$ 0.05). Clearly separable spikes  were sorted, whenever possible, by using template matching methods. The sorting quality of single neuron activity was guaranteed by the absence of short ($<$ 1\,ms) interspike intervals, the homogeneity of the waveforms included in a cluster, and their spatial segregation from those belonging to other clusters in a 2D or 3D principal components space. Furthermore, cluster stability along the recording session was guaranteed by verifying that the position of the spikes in each cluster projected in a 2D principal component space remained constant over the whole recording time \cite{Bonini2014b}. For the further comparison of spike shapes across multiple recording days the data from different sessions were merged together and spike sorting algorithms were applied with the same parameters to the whole dataset to be compared.
To assess possible LFP modulation, we applied a multitaper spectral analysis. In short, a Fourier transformation was applied to the tapered time series signal. We used an optimal family of orthogonal tapers, the prolate spheroidal (Slepian) functions that are parametrized by their time length $T$ and the frequency bandwidth $W$. For each choice of $T$ and $W$, a maximal number of $K = 2TW - 1$ tapers could be used for spectral estimation. In this study, we used $K = 5$ (i.e. $TW = 3$) and a time window of 500\,ms with a step size of 50\,ms. Data were sampled at 2\,kHz and the attention was focused on a 0-40\,Hz frequency range.

\section{Results}

Before fabrication of the cylindrical probes, the layout of recording sites had to be chosen. Based on the magnetic resonance (MR) data of the specific monkey we designed the electrode sites to target the anterior intraparietal (AIP) area and approach the mesial wall of the hemisphere, at the boundary between the mesial parietal and the posterior cingulate cortices.

\subsection{Characterization}
\label{sec:Character}
\begin{figure}[tbh]
	\centering
	\includegraphics[width=15.7cm]{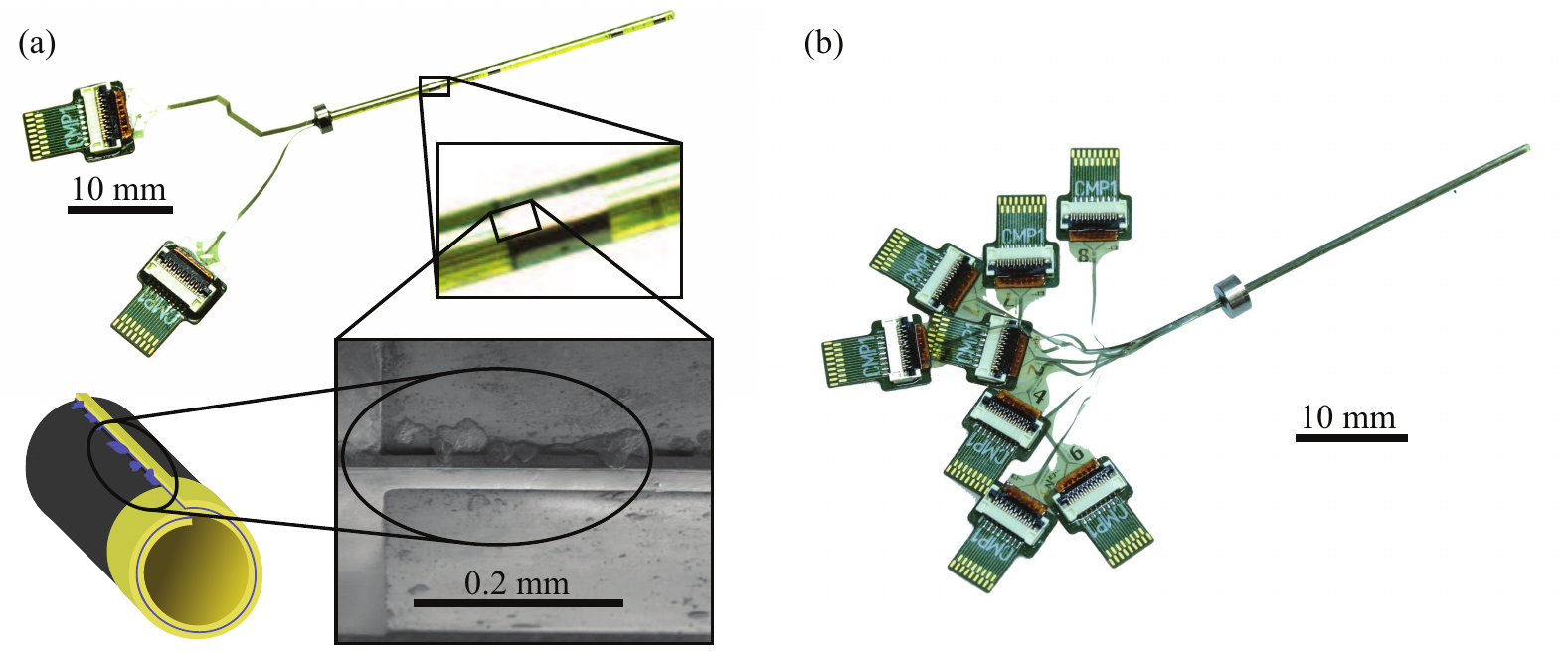}
	\caption{(a) (top) Photograph of the 32-channel probe with magnified inset and (left-bottom) schematic probe section, both indicating the position of the (right-bottom) detailed SEM view; (b)\,photograph of the 128-channel probe.}
	\label{fig:SEM}
\end{figure}

\begin{figure}[tbh]
	\centering
	\includegraphics[width=8cm]{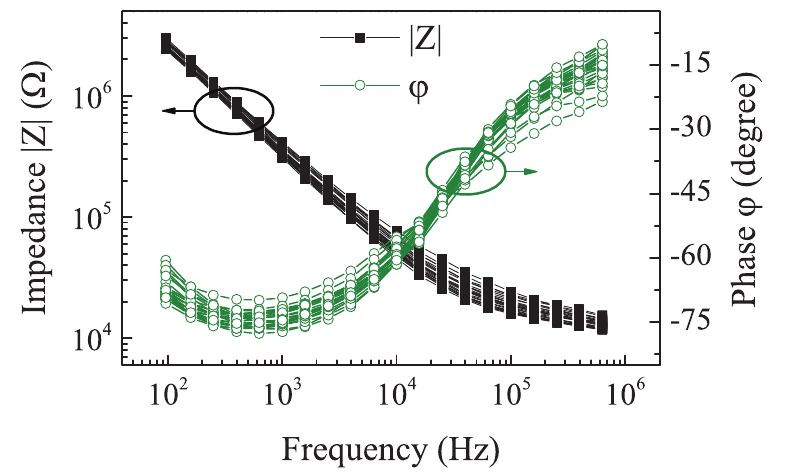}
	\caption{Absolute impedance $\left|Z\right|$ and phase $\varphi$ versus frequency of the 26 functional microelectrodes of the implanted 32-channel Probe P2.}
	\label{fig:Impedance}
\end{figure}

The calibrated optical microscope revealed that the diameters of nine different probes were 794\,$\mu$m on average with a standard deviation of 6\,$\mu$m. We successfully inserted the probes through the guiding screws and they fitted easily without noticeable play.
We performed SEM on one of the probes not used for implantation. In the magnified view in Figure\,\ref{fig:SEM}(a), the polyimide step from the top PI layer down to an electrode is easily recognized. There is no visible gap between the end of the foil and its preceding revolution, as illustrated in the lower left part of Figure\,\ref{fig:SEM}(a). In the highlighted section we observed some protrusions beyond the end of the foil. These were interpreted as excess Cytop that had leaked from between the two PI revolutions. Figure\,\ref{fig:SEM}(b) shows a 128-channel probe with eight connectors for 16 channels each.
Insertion into the agar gel brain model showed no mechanical damage nor buckling of any of the probes.
A representative impedance spectrogram of one of the implanted 32-channel probes (with 4 macroelectrodes and 28 microelectrodes) is shown in Figure\,\ref{fig:Impedance} with absolute impedance and phase values of its 26 functional microelectrodes. The absolute impedance $\left|Z\right|$ at 1\,kHz of the Pt electrodes with a diameter of 35\,$\mu$m was extracted to exhibit an average value of 354\,k$\Omega\pm$31\,k$\Omega$ standard deviation.

\begin{figure}[tbh]
	\centering
	\includegraphics[width=8cm]{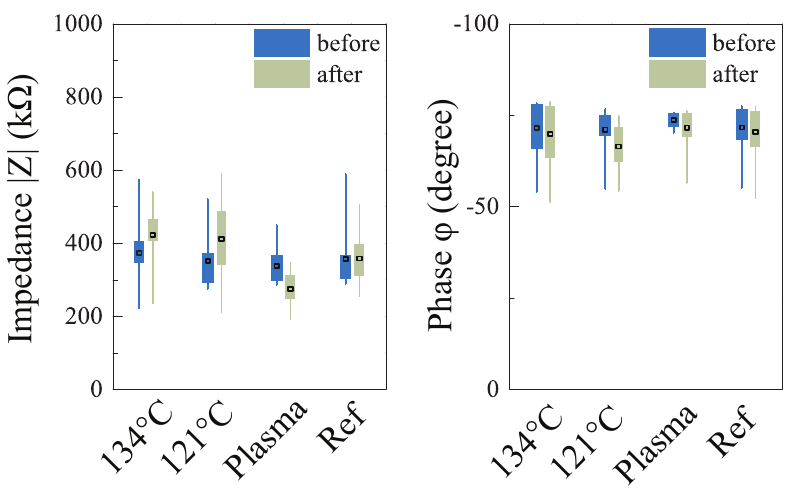}
	\caption{Impedance and phase measurements at 1\,kHz for three different sterilization procedures, i.e. autoclave at 134$^\circ$C, and at 121$^\circ$C, and a STERRAD\textsuperscript{\textregistered} plasma process. Measurements were performed on four 64-channel probes (one for each sterilization method, and one reference not undergoing sterilization). The boxes represent the interquartile ranges, while the whiskers represent the 5th/95th percentiles. Only data of functional channels are reported.}
	\label{fig:steri}
\end{figure}

Each of the probes steam sterilized at 134$^\circ$C or 121$^\circ$C exhibited 4 broken channels before the sterilization process and  7 broken channels thereafter. The STERRAD\textsuperscript{\textregistered} plasma sterilized probe showed 1 channel with an impedance above 1\,M$\Omega$ before and 1 after sterilization. 
The reference probe exhibited 3 channels without function during both measurements. The data suggest that all the tested sterilization methods have a minor impact on the channels integrity. The autoclave process seems to harm the function of at least some of the electrodes of the probe.
To better assess more subtle differences in the impact of different sterilization methods on the channels functionality, all channels with an impedance above 1\,M$\Omega$ after sterilization were disregarded from the further analysis. A box plot of the absolute impedance and phase of these channels are given in Figure\,\ref{fig:steri}.
The change in impedance of the sterilized probes appears to be low compared to the variance of the electrodes of one single probe right after fabrication. From this point of view all sterilization methods are viable candidates. However, the plasma process seems to be more benign with regard to probe functionality.

\subsection{In vivo measurements}
\label{sec:Measure}

We were able to test all three probes P1 to P3 in vivo. Their characteristics and recording performance are summarized in Table\,\ref{tab:summary} at day 1 after the implantation.

\begin{table*}[tbh]
	\centering
		\begin{tabular}{|l|c|c|c|}
			\hline
			\textbf{Identifier} & \textbf{P1} & \textbf{P2} & \textbf{P3} \\
			\hline
			Macro-/microelectrodes & 4/28 & 4/28 & -/64 \\
			\hline
			Functional microelectrodes & 24 & 26 & 62 \\
			\hline
			Probe length (mm) & 30 & 30 & 30 \\
			\hline
			Channels with LFP & 21 & 19 & * \\
			\hline
			Channels with MUA & 5 & 12 & 26 \\
			\hline
			Channels with SUA & - & 7 & 11 \\
			\hline
			Sorted single units & - & 8 & 16 \\
			\hline
		\end{tabular}
	\caption{Summary of probe characteristics, and recording results of the three implanted probes one day after implantation. *Because signals of P3 were recorded using the Open Ephys system, the offline processing work flow used for P1 and P2 could not eliminate all artefacts in the low-frequency regime for P3.}
	\label{tab:summary}
\end{table*}

For probe P1, two days after its implantation clear modulations of both the LFP and MUA were recorded during the grasping task. Because of a technical problem, the monkey was able to reach and damage the connector of P1 at the end of the recording session on day 2, so we could not collect more data from this probe.

\begin{figure}[tbh]
	\centering
	\includegraphics[width=8cm]{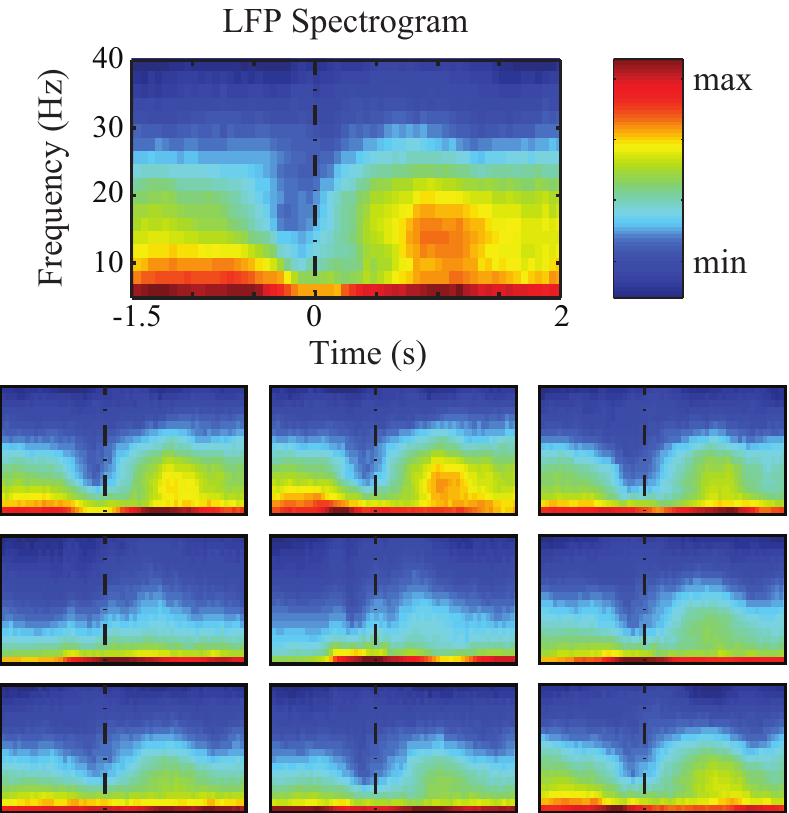}
	\caption{Peri-event spectrograms (LFPs) recorded with probe P2, 1 hour after implantation, aligned to the beginning of the object pulling phase during the execution of the grasping task (time 0). The amplitudes of the 10 channels are normalized to their respective maxima and minima.}
	\label{fig:perieventhour}
\end{figure}

\begin{figure}[tbh]
	\centering
	\includegraphics[width=8cm]{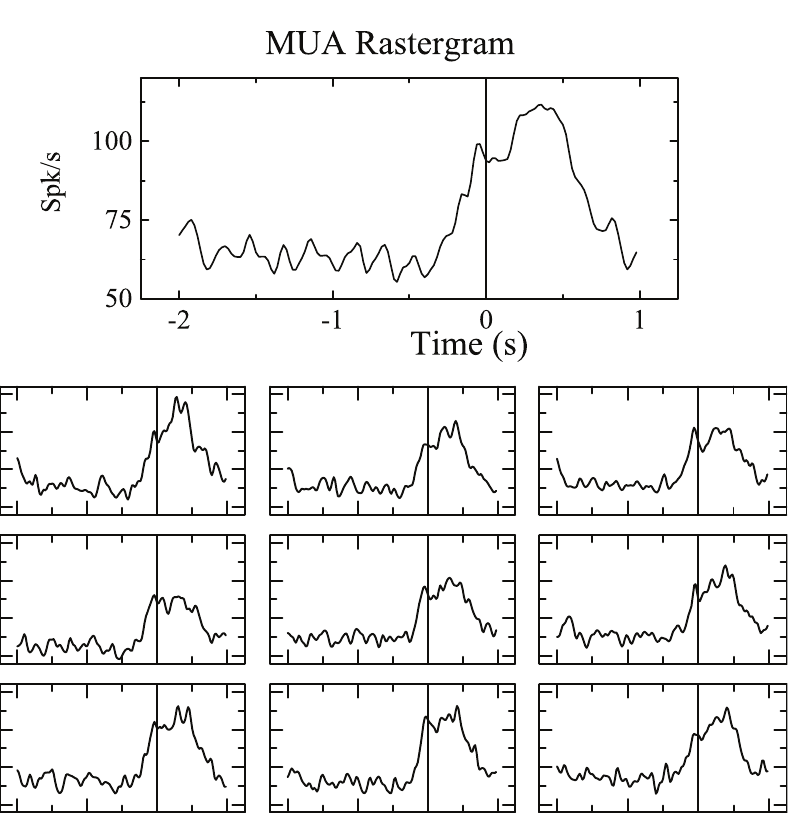}
	\caption{Time series of MUA simultaneously recorded with probe P2, 1 hour after implantation from the same channels with the same temporal alignment as in Figure\,\ref{fig:perieventhour}. Spike detection has been performed by setting the threshold to the threefold standard deviation from the mean height of the signal peak. Abscissa and ordinate scales are the same for all channels.}
	\label{fig:MUARasterhour}
\end{figure}

\begin{figure}[tbh]
	\centering
	\includegraphics[width=8cm]{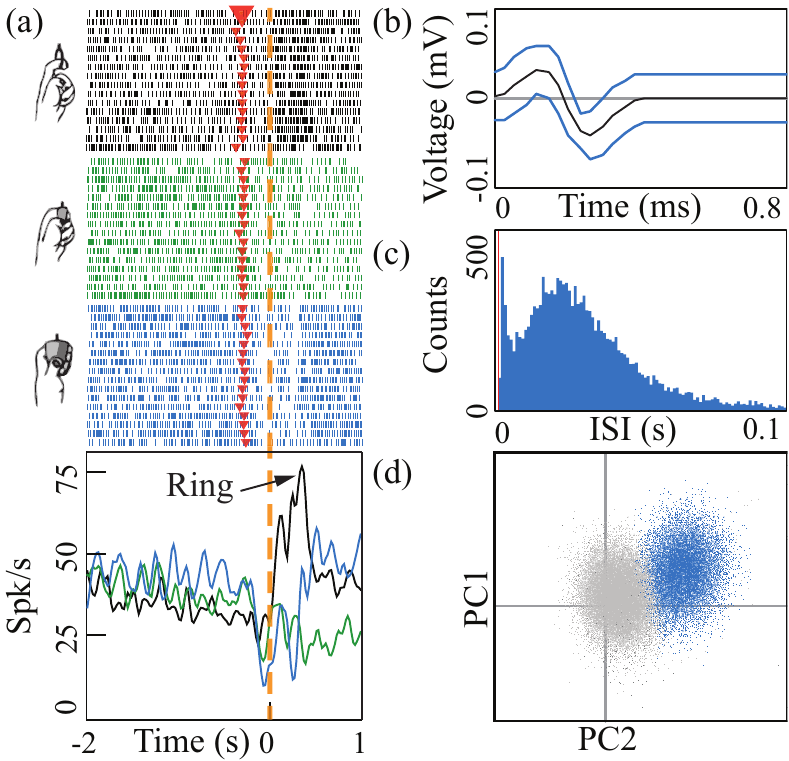}
	\caption{(a) Rastergrams and spike density functions of an isolated neuron recorded one day after implantation during the execution of the grasping motor task with three different objects aligned with respect to the pulling onset (orange dashed line). Note that this cell is tonically active, with a firing rate of approximately 40 spikes/s, and it does respond during grasping of the ring while it is inhibited during the hand shaping phase prior to grasping of the other two objects, particularly the big cone. The red triangles indicate the onset of the reaching movements. (b) Spike shape of the neuron, (c) inter-spike interval distribution of the neuron, and (d) example of two principle components (PC1 and PC2) distinguishing this neuron (blue) from the non-sorted units (grey).\newline}
	\label{fig:Unitrastergram}
\end{figure}

\begin{figure}[tbh]
	\centering
	\includegraphics[width=8cm]{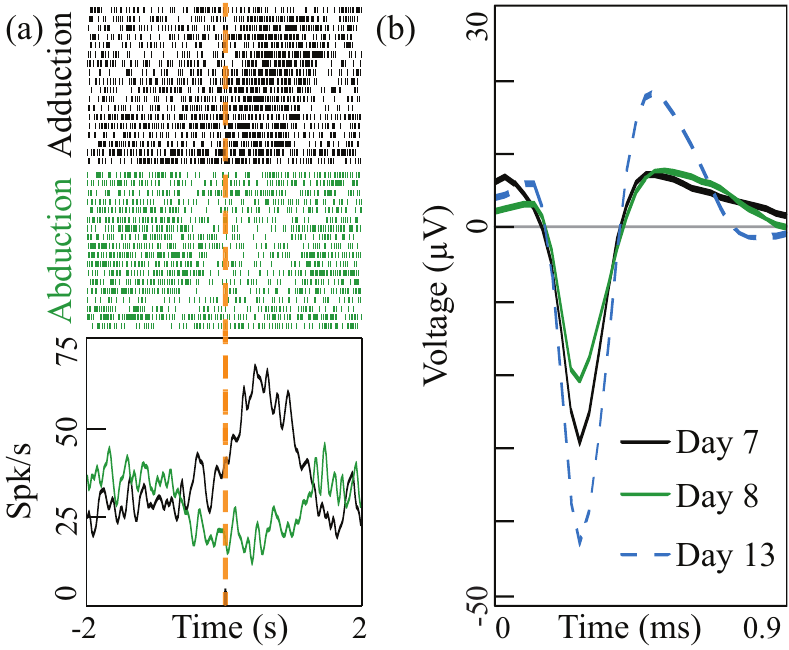}
	\caption{(a) Rastergrams and spike density functions of a single neuron recorded during adduction and abduction of the monkey’s contralateral foot aligned with respect to  the beginning of the joint mobilization (orange dashed line). (b) Spike shape and amplitude across multiple days of recording of the unit shown in (a).}
	\label{fig:foot}
\end{figure}

Probe P2 was implanted four days later. We followed the same surgical procedure as for P1: Since the probe was positioned inside the same protective chamber implanted during the first operation, the surgery was faster and required only a single dose of anaesthetics. This allowed us to test the probe functionality as early as one hour after implantation. Both LFP and MUA were detected in this first session, and SUA clearly appeared one day later (see details in Table 1 for recording results obtained with P2 on day 1). Figure\,\ref{fig:perieventhour} shows examples of the peri-event spectrograms of the low frequency bands recorded from different microelectrodes from probe P2 one hour after implantation. Spike detection of MUA was performed by setting the voltage crossing threshold to the three-fold standard deviation of the signal peak level of the corresponding electrodes \{Figure\,\ref{fig:MUARasterhour}\}, both about one hour after insertion. A modulation of MUA and LFP during the grasping task is clearly visible one hour after implantation and persists over the entire course of the recordings. In one of the channels a well-isolated single unit appeared one day after the implantation, with a grip-selective modulation of its activity during grasping of different objects \{Figure\,\ref{fig:Unitrastergram}\}. On day 7 after the implantation of probe P2 we detected a single unit responding specifically to proprioceptive stimulation of the monkey's contralateral foot \{Figure\,\ref{fig:foot}(a)\}, which probably had passed unnoticed on the previous days, and remained stable until day 13 \{Figure\,\ref{fig:foot}(b)\}. These preliminary recording experiments can be considered positive, as the probes allowed us to record single cells for up to 13 days after the implantation.

\begin{figure}[tbh]
	\centering
	\includegraphics[width=8cm]{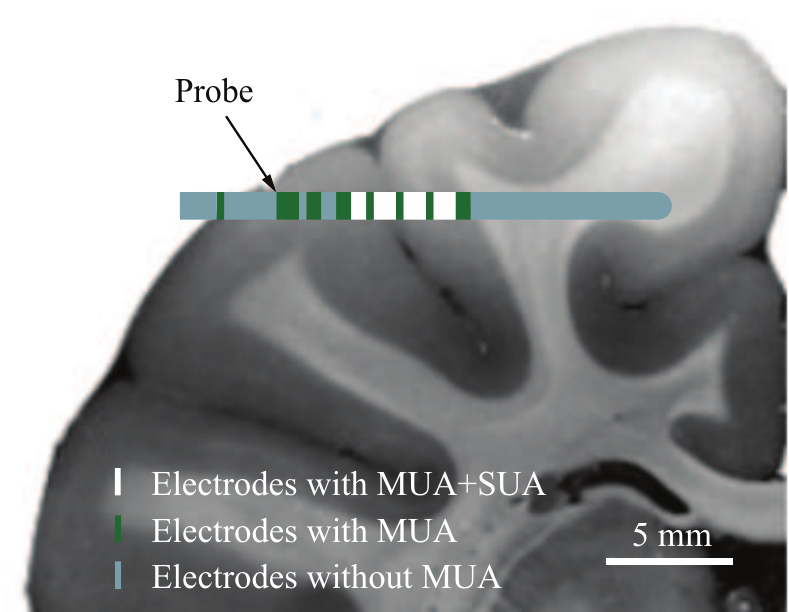}
	\caption{MR-image from "{}The Scalable Brain Atlas"{}\cite{Bakker2015} - "{}Macaque - NeuroMaps Atlas (2008)"{} with overlayed pathway of the 64-channel probe P3 at the lateral and medial walls of the intraparietal sulcus, including area AIP and the superior parietal cortex. The overlay indicates the position of electrodes recording SUA, MUA or LFPs only. From a total of 11 electrodes, 16 single units could be sorted (indicated in white). Note that the pathway has been increased in width for visibility reasons.}
	\label{fig:trajectory}
\end{figure}

After having analysed the results from P2, we fabricated and implanted the 64-channel probe P3 inside the same chamber. Not all microelectrodes were able to pick up SUA or MUA most likely due to their positioning in the white matter. For illustration, Figure\,\ref{fig:trajectory} shows the estimated trajectory of probe P3 and the actual yield of SUA and MUA, obtained at different positions along the shaft.

\begin{figure}[tbh]
	\centering
	\includegraphics[width=15.7cm]{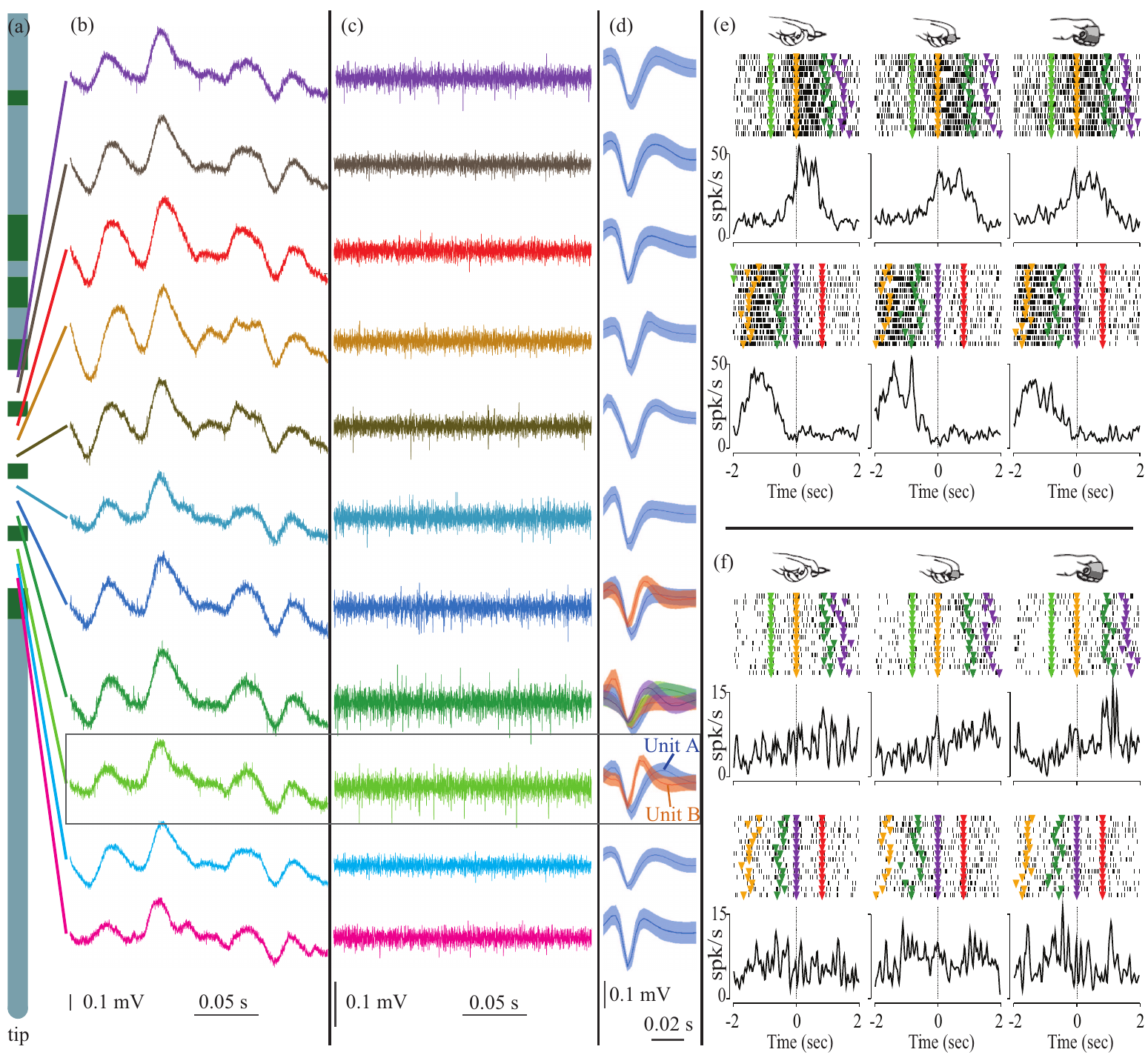}
	\caption{(a) Schematic probe with electrode positions in accordance with Figure\,\ref{fig:trajectory} illustrating the position of (b) exemplary 0.2\,second wideband signals of the 11 channels from the 64-channel probe P3 showing identifiable SUA. (c) Same data as in (b) filtered with a fourth-order 300 Hz Bessel low-cut filter. (d) Single-unit wave forms ($\pm$1 standard deviation) have been sorted from the whole acquisition period (710\,s) from which the traces shown in (c) have been taken. (e) Rastergrams and histograms of the activity of Unit A (blue) and (f) Unit B (orange) isolated from the same channel (highlighted signal trace) during visual presentation (upper halves of (e) and (f)) and pulling (lower halves) of the three objects represented on top of each column. Coloured markers indicate: object presentation (orange); onset/offset of the cue sound (light green/dark green); object pulling onset (purple); reward (red).}
	\label{fig:64ch}
\end{figure}

Probe P3 showed signals of similar quality in the high frequency bands as the implanted 32-channel probes. Representative data from the 11 electrodes showing SUA are given in Figure\,\ref{fig:64ch}; the respective position of these electrodes is indicated in the schematic probe representation in Figure\,\ref{fig:64ch}(a). Exemplary 0.2\,s long wideband traces of these 11 electrodes Bessel high-pass filtered at 300\,Hz are depicted in Figure\,\ref{fig:64ch}(c) to visualize the spike trains. Figure\,\ref{fig:64ch}(d) shows the corresponding SUA sorted from the filtered signals. Figure\,\ref{fig:64ch}(e) shows six rastergrams of the activity of Unit A (blue) extracted from signal trace 9 from the top for the three different objects and the grasping task presented in Figure\,\ref{fig:Unitrastergram}. The upper set of rastergrams and spike density plots were aligned to the object presentation phase of the task while the lower set shows the same data aligned on the object pulling phase. Figure\,\ref{fig:64ch}(f) shows the same rastergrams as in Figure\,\ref{fig:64ch}(e) from Unit B (orange). In contrast to Unit A, Unit B is not modulated by the visuomotor task.

\begin{figure}[tbh]
	\centering
	\includegraphics[width=8cm]{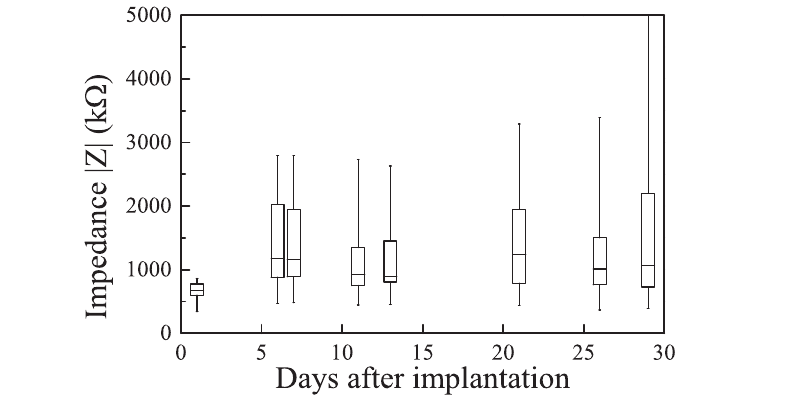}
	\caption{Median (center line),  interquartile range (box), and 5th/95th percentile (whiskers) of the absolute value of electrode impedance at 1\,kHz as a function of time for 58 microelectrodes of probe P3 that exhibited an impedance value below 1\,M$\Omega$ on Day 1 after the implantation. In total 6 microelectrodes were omitted from this analysis as they showed an impedance above 1\,M$\Omega$ after the implantation.}
	\label{fig:imp}
\end{figure}

Although the signal quality decreased over time, SUA could still be recorded 26 days after implantation of the 64-channel probe. On day 29 after probe implantation all SUA had disappeared. This observation fits to the pronounced increase of the electrode impedances measured in vivo over time, as documented in Figure\,\ref{fig:imp}. Nonetheless, recording of SUA over the course of 26 days is well beyond the time needed for epileptic focus detection which typically requires around 10 days.

The 128-channel prototype shown in Figure\,\ref{fig:SEM}(b) has been characterized and shows similar impedance results as probes P1 to P3, but has not yet been implanted, as a faster and more stable interconnection approach for this large number of channels is currently being developed to facilitate probe implantation and use.

\section{Conclusion}

We have presented the fabrication and characterization of cylindrical multielectrode probes that can record physiologically meaningful data already one hour after implantation. 
The selected materials used for the probe realization have shown good results regarding the detection of SUA over a long time.
The design of the probe/electrode positions can easily be tailored for each individual subject.

The presented probe is suitable for high-channel-count depth recordings, regardless of the size or shape of the electrode sites. Small electrode sites can pick up LFPs, MUAs as well as SUAs. Thus, using this device in epileptic patients may dramatically improve the precision of epileptic focus detection. 

Truccolo et al. have measured single neuron activity during an epileptic seizure in the cortex of human patients \cite{Truccolo2011}. Other groups have measured single neurons deep within the cortex during an epileptic seizure of epileptic patients and LFPs along the shank of the SEEG probe \cite{Alvarado2013}, using micro-wires protruding from the tip of cylindrical SEEG probes and macroelectrodes along the probe shank respectively. In addition, Worrell et al. have presented high-frequency recordings along the shank of cylindrical probes \cite{Worrell2008}. These recordings, including SUA during epileptic seizures have led to a more detailed understanding of focal epilepsy. The device described in this work not only increases the resolution of the epileptic focus localization but could further enable single neuron activity recordings along the entire shank of the probe at high-resolution, leading to an even more refined electrical image of focal epilepsy.

The electrode sites have not been characterized for the effectiveness of electrical stimulation of neural tissue yet, but the manufacturing process and design in principle allow for stimulation. Moreover, the space within the PI cylinder can further be be taken advantage of for various purposes such as optogenetic stimulation of neural tissue as proposed in \cite{Schwaerzle2015}. Thus neural dysfunctions beyond focal epilepsy may also be addressed using the novel device. In applications of this device exceeding a few weeks of implantation, further studies have to identify potential causes of failure of the probe and materials. These studies include suitable technical solutions to overcome long-term signal degradation. Studies on biocompatibility are currently ongoing.

\ack
The research leading to these results has received funding from the European Union's Seventh Framework Program (FP7/2007-2013) under grant agreement n$^\circ$600925.
The authors gratefully acknowledge the support from  Juan Ordonez for discussions on PI process technologies, Dennis Plachta for providing us with the first needle to roll the PI foils, Andreas Gehringer for fabricating our rolling needles, IMTEK's Clean Room Service Center staff, and the support from Carolin Gierschner for helping us with the sterilization tests.

\section*{References}
\bibliographystyle{unsrt}

\bibliography{ref}

\begin{thebibliography}{10}

\bibitem{WHO2006}
WHO.
\newblock {\em Neurological disorders public health challenges}.
\newblock World Health Organization, Geneva, 2006.

\bibitem{Scarabin2012}
J.M. Scarabin.
\newblock {\em Stereotaxy and Epilepsy Surgery - With videos (English and
  French Edition)}.
\newblock John Libbey Eurotext ltd, 2012.

\bibitem{Cossu2005}
M.~Cossu, F.~Cardinale, L.~Castana, A.~Citterio, S.~Francione, L.~Tassi, A.L.
  Benabid, and G.~Lo~Russo.
\newblock Stereoelectroencephalography in the presurgical evaluation of focal
  epilepsy: A retrospective analysis of 215 procedures.
\newblock {\em Neurosurgery}, 57(4):706--718, 2005.

\bibitem{Rodriguez2005}
M.C. Rodriguez-Oroz, J.A. Obeso, A.E. Lang, J.-L. Houeto, P.~Pollak,
  S.~Rehncrona, J.~Kulisevsky, A.~Albanese, J.~Volkmann, M.I. Hariz, N.P.
  Quinn, J.D. Speelman, J.~Guridi, I.~Zamarbide, A.~Gironell, J.~Molet,
  B.~Pascual-Sedano, B.~Pidoux, A.M. Bonnet, Y.~Agid, J.~Xie, A.-L. Benabid,
  A.M. Lozano, J.~Saint-Cyr, L.~Romito, M.F. Contarino, M.~Scerrati, V.~Fraix,
  and N.~Van~Blercom.
\newblock Bilateral deep brain stimulation in parkinson{\textquoteright}s
  disease: {A} multicentre study with 4 years follow-up.
\newblock {\em Brain}, 128(10):2240--2249, 2005.

\bibitem{Wishart2003}
H.A. Wishart, D.W. Roberts, R.M. Roth, B.C. McDonald, D.J. Coffey, A.C.
  Mamourian, C.~Hartley, L.A. Flashman, C.E. Fadul, and A.J. Saykin.
\newblock Chronic deep brain stimulation for the treatment of tremor in
  multiple sclerosis: Review and case reports.
\newblock {\em J. Neurol., Neurosurgery \& Psychiatry}, 74(10):1392--1397,
  2003.

\bibitem{Mayberg2005}
H.S. Mayberg, A.M. Lozano, V.~Voon, H.E. McNeely, D.~Seminowicz, C.~Hamani,
  J.M. Schwalb, and S.H. Kennedy.
\newblock Deep brain stimulation for treatment-resistant depression.
\newblock {\em Neuron}, 45(5):651 -- 660, 2005.

\bibitem{Houeto2005}
J.L. Houeto, C.~Karachi, L.~Mallet, B.~Pillon, J.~Yelnik, V.~Mesnage, M.L.
  Welter, S.~Navarro, A.~Pelissolo, P.~Damier, B.~Pidoux, D.~Dormont, P.~Cornu,
  and Y.~Agid.
\newblock Tourette’s syndrome and deep brain stimulation.
\newblock {\em J. Neurol., Neurosurgery \& Psychiatry}, 76(7):992--995, 2005.

\bibitem{Picot2008}
M.C. Picot, M.~Baldy-Moulinier, J.P. Daur{\`e}s, P.~Dujols, and A.~Crespel.
\newblock The prevalence of epilepsy and pharmacoresistant epilepsy in adults:
  A population-based study in a western european country.
\newblock {\em Epilepsia}, 49(7):1230--1238, 2008.

\bibitem{Truccolo2011}
W.~Truccolo, J.A. Donoghue, L.R. Hochberg, E.N. Eskandar, J.R. Madsen, W.S.
  Anderson, E.N. Brown, E.~Halgren, and S.S. Cash.
\newblock Single-neuron dynamics in human focal epilepsy.
\newblock {\em Nature Neurosci.}, 14(5):635--641, May 2011.

\bibitem{Schulze2014}
A.~Schulze-Bonhage and J.~Zentner.
\newblock The preoperative evaluation and surgical treatment of epilepsy.
\newblock {\em Deutsches {\"A}rzteblatt International}, 111(18):313, 2014.

\bibitem{Avanzini2016}
P.~Avanzini, R.O. Abdollahi, I.~Sartori, F.~Caruana, V.~Pelliccia, G.~Casaceli,
  R.~Mai, G.~Lo~Russo, G.~Rizzolatti, and G.A. Orban.
\newblock Four-dimensional maps of the human somatosensory system.
\newblock {\em Proceedings of the National Academy of Sciences}, page
  201601889, 2016.

\bibitem{Misra2014}
A.~Misra, J.F. Burke, A.G. Ramayya, J.~Jacobs, M.R. Sperling, K.A. Moxon, M.J.
  Kahana, J.J. Evans, and A.D. Sharan.
\newblock Methods for implantation of micro-wire bundles and optimization of
  single/multi-unit recordings from human mesial temporal lobe.
\newblock {\em J. Neural Eng.}, 11(2):026013, 2014.

\bibitem{Worrell2008}
G.A. Worrell, A.B. Gardner, S.M. Stead, S.~Hu, S.~Goerss, G.J. Cascino, F.B.
  Meyer, R.~Marsh, and B.~Litt.
\newblock High-frequency oscillations in human temporal lobe: {S}imultaneous
  microwire and clinical macroelectrode recordings.
\newblock {\em Brain}, 131(4):928--937, 2008.

\bibitem{Fried1997}
I.~Fried, K.A. MacDonald, and C.L. Wilson.
\newblock Single neuron activity in human hippocampus and amygdala during
  recognition of faces and objects.
\newblock {\em Neuron}, 18(5):753--765, 1997.

\bibitem{Kawasaki2001}
H.~Kawasaki, R.~Adolphs, O.~Kaufman, H.~Damasio, A.R. Damasio, M.~Granner,
  H.~Bakken, T.~Hori, and M.A. Howard.
\newblock Single-neuron responses to emotional visual stimuli recorded in human
  ventral prefrontal cortex.
\newblock {\em Nature Neurosci.}, 4(1):15--16, 2001.

\bibitem{Mukamel2010}
R.~Mukamel, A.D. Ekstrom, J.~Kaplan, M.~Iacoboni, and I.~Fried.
\newblock Single-neuron responses in humans during execution and observation of
  actions.
\newblock {\em Curr. Biol.}, 20(8):750--756, 2010.

\bibitem{Hefft2013}
S.~Hefft, A.~Brandt, S.~Zwick, D.~von Elverfeldt, I.~Mader, J.~Cordeiro,
  M.~Trippel, J.~Blumberg, and A.~Schulze-Bonhage.
\newblock Safety of hybrid electrodes for single-neuron recordings in humans.
\newblock {\em Neurosurgery}, 73(1):78--85, 2013.

\bibitem{Alvarado2013}
C.~Alvarado-Rojas, K.~Lehongre, J.~Bagdasaryan, A.l Bragin, R.~Staba,
  J.~Engel~Jr, V.~Navarro, and M.~Le~Van~Quyen.
\newblock Single-unit activities during epileptic discharges in the human
  hippocampal formation.
\newblock {\em Frontiers in Computational Neuroscience}, 7, 2013.

\bibitem{Gompel2008}
J.J. Van~Gompel, S.M. Stead, C.~Giannini, F.B. Meyer, W.R. Marsh, T.~Fountain,
  E.~So, A.~Cohen-Gadol, K.H. Lee, and G.A. Worrell.
\newblock Phase {I} trial: safety and feasibility of intracranial
  electroencephalography using hybrid subdural electrodes containing macro-and
  microelectrode arrays.
\newblock {\em Neurosurgical focus}, 25(3):E23, 2008.

\bibitem{Ulbert2004}
I.~Ulbert, Z.~Magl\'{o}czky, L.~Er\H{o}ss, S.~Czirj\'{a}k, J.~Vajda,
  L.~Bogn\'{a}r, S.~T\'{o}th, Z.~Szab\'{o}, P.~Hal\'{a}sz, D.~Fab\'{o},
  E.~Halgren, T.F. Freund, and G.~Karmos.
\newblock In vivo laminar electrophysiology co-registered with histology in the
  hippocampus of patients with temporal lobe epilepsy.
\newblock {\em Experimental Neurology}, 187(2):310 -- 318, 2004.

\bibitem{Csercsa2010}
R.~Csercsa, B.~Dombov{\'a}ri, D.~Fab{\'o}, L.~Wittner, L.~Er{\H{o}}ss, L.~Entz,
  A.~S{\'o}lyom, G.~R{\'a}sonyi, A.~Sz{\H{u}}cs, A.~Kelemen, R.~Jakus,
  V.~Juhos, L.~Grand, A.~Magony, P.~Hal{\'a}sz, T.F. Freund, Z.~Magl{\'o}czky,
  S.S. Cash, L.~Papp, G.~Karmos, E.~Halgren, and I.~Ulbert.
\newblock Laminar analysis of slow wave activity in humans.
\newblock {\em Brain}, page awq169, 2010.

\bibitem{Fomani2011a}
A.A. Fomani, R.R. Mansour, C.M. Florez-Quenguan, and P.L. Carlen.
\newblock Development and characterization of multisite three-dimensional
  microprobes for deep brain stimulation and recording.
\newblock {\em J. Microelectromech. Systems}, 20(5):1109--1118, 2011.

\bibitem{Puije1989}
P.D. van~der Puije, C.R. Pon, and H.~Robillard.
\newblock Cylindrical cochlear electrode array for use in humans.
\newblock {\em The Ann. of Otology, Rhinology, and Laryngology},
  98(6):466--471, 1989.

\bibitem{Mercanzini2010}
A.~Mercanzini and P.~Renaud.
\newblock Microfabricated neurostimulation device, {P}atent number
  {W}{O}2010055421{A}1, May 2010.

\bibitem{Martens2008}
H.C.F. Martens, M.M.J. Decré, and E.~Cantatore.
\newblock Electrode system for deep brain stimulation, {P}atent number
  {W}{O}2008107815{A}1, September 2008.

\bibitem{Hetke1990}
J.F. Hetke, K.~Najafi, and K.D. Wise.
\newblock Flexible miniature ribbon cables for long-term connection to
  implantable sensors.
\newblock {\em Sensors and Actuators A: Physical}, 23(1â€“3):999 -- 1002,
  1990.
\newblock Proceedings of the 5th International Conference on Solid-State
  Sensors and Actuators and Eurosensors \{III\}.

\bibitem{Stieglitz2000}
T.~Stieglitz, H.~Beutel, M.~Schuettler, and J.-U. Meyer.
\newblock Micromachined, polyimide-based devices for flexible neural
  interfaces.
\newblock {\em Biomedical Microdevices}, 2(4):283--294, 2000.

\bibitem{Rodriguez2000}
F.J. Rodriguez, D.~Ceballos, M.~Schuettler, A.~Valero, E.~Valderrama,
  T.~Stieglitz, and X.~Navarro.
\newblock Polyimide cuff electrodes for peripheral nerve stimulation.
\newblock {\em J. Neurosci. Methods}, 98(2):105 -- 118, 2000.

\bibitem{Plachta2014}
D.T.T. Plachta, M.~Gierthmuehlen, O.~Cota, N.~Espinosa, F.~Boeser, T.C.
  Herrera, T.~Stieglitz, and J.~Zentner.
\newblock Blood pressure control with selective vagal nerve stimulation and
  minimal side effects.
\newblock {\em J. Neural Eng.}, 11(3):036011, 2014.

\bibitem{Pothof2014}
F.~Pothof, S.~Anees, J.~Leupold, L.~Bonini, O.~Paul, G.A. Orban, and P.~Ruther.
\newblock Fabrication and characterization of a high-resolution neural probe
  for stereoelectroencephalography and single neuron recording.
\newblock In {\em Engineering in Medicine and Biology Society (EMBC), 2014 36th
  Annual International Conference of the IEEE}, pages 5244--5247, 2014.

\bibitem{Kisban2007}
S.~Kisban, S.~Herwik, K.~Seidl, B.~Rubehn, O.~Paul, P.~Ruther, T.~Stieglitz,
  and A.~Jezzini.
\newblock Microprobe array with low impedance electrodes and highly flexible
  polyimide cables for acute neural recording.
\newblock In {\em Engineering in Medicine and Biology Society, 2007. EMBS 2007.
  29th Annual International Conference of the IEEE}, pages 175--178. IEEE,
  2007.

\bibitem{Herwik2009}
S.~Herwik, S.~Kisban, A.A.A. Aarts, K.~Seidl, G.~Girardeau, K.~Benchenane, M.B.
  Zugaro, S.I. Wiener, O.~Paul, H.P. Neves, and P.~Ruther.
\newblock Fabrication technology for silicon-based microprobe arrays used in
  acute and sub-chronic neural recording.
\newblock {\em J. Micromech. Microeng.}, 19(7):074008, 2009.

\bibitem{Seidl2011}
K.~Seidl, S.~Herwik, T.~Torfs, H.P. Neves, O.~Paul, and P.~Ruther.
\newblock {CMOS}-based high-density silicon microprobe arrays for electronic
  depth control in intracortical neural recording.
\newblock {\em J. Microelectromech. Syst.}, 20(6):1439--1448, 2011.

\bibitem{Rubehn2009}
B.~Rubehn, C.~Bosman, R.~Oostenveld, P.~Fries, and T.~Stieglitz.
\newblock A {MEMS}-based flexible multichannel {ECoG}-electrode array.
\newblock {\em J. Neural Eng.}, 6(3):036003, 2009.

\bibitem{Schwaerzle2016}
M.~Schwaerzle, J.~Nehlich, S.~Ayub, O.~Paul, and P.~Ruther.
\newblock {LED}-based optical cochlear implant on highly flexible triple layer
  polyimide substrates.
\newblock In {\em Micro Electro Mechanical Systems (MEMS), 2016 29th IEEE
  International Conference on}, pages 395--398. IEEE, 2016.

\bibitem{Ordonez2012}
J.S. Ordonez, C.~Boehler, M.~Schuettler, and T.~Stieglitz.
\newblock Long-term adhesion studies of polyimide to inorganic and metallic
  layers.
\newblock In {\em Symposium TT – Interfaces in Materials, Biol. and Phys.},
  volume 1466 of {\em MRS Proceedings}, 2012.

\bibitem{Bonini2014}
L.~Bonini, M.~Maranesi, A.~Livi, L.~Fogassi, and G.~Rizzolatti.
\newblock Space-dependent representation of objects and other's action in
  monkey ventral premotor grasping neurons.
\newblock {\em J. Neurosci.}, 34(11):4108--4119, 2014.

\bibitem{Rozzi2008}
S.~Rozzi, P.F. Ferrari, L.~Bonini, G.~Rizzolatti, and L.~Fogassi.
\newblock Functional organization of inferior parietal lobule convexity in the
  macaque monkey: electrophysiological characterization of motor, sensory and
  mirror responses and their correlation with cytoarchitectonic areas.
\newblock {\em Eur. J. Neurosci.}, 28(8):1569--1588, 2008.

\bibitem{Maranesi2012}
M.~Maranesi, F.~Roda, L.~Bonini, S.~Rozzi, P.F. Ferrari, L.~Fogassi, and
  G.~Coud{\'e}.
\newblock Anatomo-functional organization of the ventral primary motor and
  premotor cortex in the macaque monkey.
\newblock {\em Eur. J. Neurosci.}, 36(10):3376--3387, 2012.

\bibitem{Bruni2015}
S.~Bruni, V.~Giorgetti, L.~Fogassi, and L.~Bonini.
\newblock Multimodal encoding of goal-directed actions in monkey ventral
  premotor grasping neurons.
\newblock {\em Cerebral Cortex}, page bhv246, 2015.

\bibitem{Bonini2014b}
L.~Bonini, M.~Maranesi, A.~Livi, S.~Bruni, L.~Fogassi, T.~Holzhammer, O.~Paul,
  and P.~Ruther.
\newblock Application of floating silicon-based linear multielectrode arrays
  for acute recording of single neuron activity in awake behaving monkeys.
\newblock {\em Biomedical Engineering/Biomedizinische Technik}, 59(4):273--281,
  2014.

\bibitem{Bakker2015}
R.~Bakker, P.~Tiesinga, and R.~K\"otter.
\newblock The scalable brain atlas: Instant web-based access to public brain
  atlases and related content.
\newblock {\em Neuroinformatics}, 13(3):353--366, 2015.

\bibitem{Schwaerzle2015}
M.~Schwaerzle, F.~Pothof, O.~Paul, and P.~Ruther.
\newblock High-resolution neural depth-probe with integrated 460 nm light
  emitting diode for optogenetic applications.
\newblock In {\em Dig. Tech. Papers Solid-State Sensors, 2015 18th
  International Conference on Actuators and Microsystems (TRANSDUCERS)}, 2015.

\end{thebibliography}

\end{document}